\documentclass[12pt,preprint]{aastex}

\shorttitle{Bowen Fluorescence in X1822-371}
\shortauthors{J. Casares et al.}

\begin{document}

\title{Bowen Fluorescence from the Companion Star in X1822-371}

\author{J. Casares}
\affil{Instituto de Astrof\'\i{}sica de Canarias, 38200 La Laguna, 
Tenerife, Spain}
\email{jcv@ll.iac.es}

\author{D. Steeghs\altaffilmark{1}}
\affil{Harvard-Smithsonian Center for Astrophysics, 60 Garden Street, 
MS-67, Cambridge, MA 02138}
\email{dsteeghs@cfa.harvard.edu}

\author{R.I. Hynes\altaffilmark{1,2}}
\affil{The University of Texas at Austin, Astronomy Department, 1 
University Station C1400, Austin, Texas 78712}
\email{rih@obelix.as.utexas.edu}

\author{P.A. Charles}
\affil{Dept of Physics \& Astronomy, University of Southampton, 
Southampton, SO17 1BJ, UK}
\email{pac@astro.soton.ac.uk}

\and

\author{K. O'Brien\altaffilmark{3}}
\affil{European Southern Observatory, Casilla 19001, Santiago 19, 
Chile}
\email{kobrien@eso.org}

\altaffiltext{1}{Dept of Physics \& Astronomy, University of Southampton, 
Southampton, SO17 1BJ, UK}
\altaffiltext{2}{Hubble Fellow}  
\altaffiltext{3}{School of Physics and Astronomy, University of St Andrews, 
St Andrews KY16 9SS, UK}  
 
\begin{abstract}

We present a clear evidence for the motion of the companion star in the Low 
Mass X-Ray Binary (LMXB) X1822-371. We detect NIII $\lambda$4640 
emission moving in antiphase with the radial velocity curve of the 
neutron star and produced on the X-ray heated hemisphere of the donor 
star. From the motion of this feature we derive a lower 
limit to the radial velocity semi-amplitude of the companion star $K_2 
\ge 300 \pm 8$ km s$^{-1}$, which, combined with a previous determination 
of the inclination angle and the pulsar's radial velocity curve,
yield $M_2 \ge 0.36(2) M_{\odot}$ and $M_1 \ge 1.14(6) M_{\odot}$. 
The HeI $\lambda$4471 absorption line moves at lower velocities 
($\simeq$ 225 km s$^{-1}$) and with a -0.05 phase shift,  
suggests a likely origin on the gas stream near the $L_1$ point. 
In addition, we detect an S-wave emission of OVI 
$\lambda$3811 produced by illumination of the hot-spot 
bulge by the central source. The Balmer lines are dominated by broad 
absorptions probably due to obscuration of the accretion disc 
by vertically extended cool material from the splash region and 
overflowing stream. We also derive a more accurate, and significantly 
different (compared to earlier work) systemic velocity of 
$\gamma=-44 \pm 5$ km s$^{-1}$ based on the motion of the HeII 
$\lambda$4686 wings and Doppler tomography. 
This work confirms the power of imaging the companion stars in LMXBs and 
outbursting transients using the Bowen fluorescence transitions. 

\end{abstract}


\section{Introduction}

X1822-371 is one of the brightest Low Mass X-ray Binaries 
(LMXBs) in the optical but comparatively weak in X-rays (i.e. 
$L_X/L_{opt} = 20$ cf. it is usually $\sim$ 500-1000, see van 
Paradijs \& McClintock 1995) which makes it the prototypical Accretion 
Disc Corona (ADC) source. In ADC sources, the combination of a 
high inclination and thick disc obscures the central source, 
and only X-rays scattered from material above and below the disc 
(a ``corona'') can reach the observer (White \& Holt 1982). 
The optical light curve displays regular eclipses of 
the accretion disc by the companion star on the 5.57 hr orbital 
period (see e.g. Hellier \& Mason 1989). Detailed modelling of the optical 
and X-ray light 
curves has provided an accurate determination of the inclination 
angle $i=82.5^{\circ} \pm 1.5^{\circ}$ (Heinz \& Nowak 2001; see 
also Hellier \& Mason 1989). 

X1822-371 is also remarkable thanks to the discovery of 0.59s X-ray 
pulses (Jonker \& van der Klis 2001, hereafter JvdK) which flags it as 
one of the ``rare" precursors of millisecond pulsars (eight out of several 
hundred 
LMXBs). The analysis of the pulse arrival time delay has enabled 
an extremely precise determination of the orbit of the neutron 
star ($a \sin i = 1.006$ lightseconds) corresponding to a primary 
radial velocity semi-amplitude of 94.5 $\pm$ 0.5 km s$^{-1}$.    
 
A complete determination of the system parameters however 
requires the knowledge of the radial velocity curve of the 
companion star. This was first intimated by Harlaftis, Charles \& Horne 
(1997) from the motion of HeI $\lambda$5876 absorption, 
which was interpreted as being produced on the irradiated face of the 
companion star, thereby providing a lower limit to the velocity 
semi-amplitude of $K_2 \ge 225 \pm 23$ km s$^{-1}$. More recently, 
Jonker, van der Klis \& Groot (2002) report $K_2=327 \pm 17$ km s$^{-1}$ 
based on measurements of HeI $\lambda$4026 and $\lambda$5875. However, 
they also find a -0.08 phase shift with respect to the pulse-timing ephemeris 
which they attribute to asymmetric irradiation of the companion's Roche 
lobe and a systemic velocity of 67 $\pm$ 15 km s$^{-1}$, 
significantly larger than in Harlaftis et al. (19 $\pm$ 19 km s$^{-1}$).   

In this paper we present new phase-resolved intermediate resolution 
blue spectroscopy of X1822-371. Inspired by our work 
on the LMXB Sco X-1 (Steeghs \& Casares 2002, herefater SC) we are pursuing the detection 
of the companion star via Bowen fluorescence emission due to X-ray irradiation. 
The combination of velocity information and light curves enable us to 
locate the site of the main emission/absorption lines whithin the binary. 

\section{Observations and Data Reduction}

We observed X1822-371 using the RGO Spectrograph 
attached to the 3.9m Anglo-Australian Telescope (AAT) 
on the nights of 10-11 June 2002. A total of 38 spectra were obtained 
with the R1200B grating centered at 4350 \AA~  which provided a 
wavelength coverage of $\lambda\lambda$3500-5250, although the extreme 
150 \AA~ at each end were severely affected by vignetting and not employed 
in the analysis.
The seeing was variable ($1.1^{\arcsec}-2^{\arcsec}$) during our run and 
we used a 
%150$\mu$ 
1 arcsec slit which resulted in a resolution of 70 km 
s$^{-1}$ (FWHM). The flux standard Feige 15 was observed with the same
instrumental configuration so as to correct for the instrumental response 
of the detector. We also rotated the slit to include a comparison 
star which enabled us to monitor slit losses and obtain a relative flux 
calibration. In addition, we obtained five more spectra of X1822-371 
with EMMI and grating $No~6$ on ESO'S 3.5m New Technology Telescope 
(NTT) at La Silla on 11 June 2002. These spectra cover the 
spectral range $\lambda\lambda$4400-5150 at 75 km s$^{-1}$ resolution. 
In total we covered 1.5 orbital cycles of the target on June 10 and one 
cycle on June 11. 

The images were de-biased and flat-fielded, and the spectra subsequently 
extracted using conventional optimal extraction techniques in order to  
optimize the signal-to-noise ratio of the output (Horne 1986).  
CuAr comparison lamp images were obtained every 30 mins, and the 
$\lambda$-pixel scale was derived through 5th-order polynomial fits to 
95 lines, resulting in an rms scatter $<$ 0.025 \AA. The calibration 
curves were interpolated linearly in time. 

\section{Trailed Spectra and Radial Velocity Curves}

Absolute flux calibration of X1822-371 and its slit comparison was tied 
to the spectrophotometric standard Feige 15. 
Slit losses for individual 
spectra were computed by fitting a second-order spline to the ratio 
between each comparison star spectrum and its grand sum average. The 
X1822-371 spectra were subsequently divided by their corresponding 
splines in order to get the final fluxed spectra, corrected for slit losses. 
Figure 1 shows the average spectra of X1822-371 and its slit comparison 
in mJy flux units. The spectrum is dominated by high excitation emission 
lines of HeII $\lambda$4686 and the Bowen blend. There is also evidence of 
a broad emission feature at $\simeq \lambda$3813 which 
was already noted by Charles, Thorstensen \& Barr (1980) in their low 
resolution spectra and tentatively identified with either HeII 
$\lambda$3813 or OVI $\lambda$3811 (see however Mason et al. 1982 for a 
different interpretation).   
%we have tentatively identified with OVI $\lambda$3811.4. 
The HeI and Balmer lines are dominated by broad 
absorptions, with residual emission cores observable in H$\beta$ and 
H$\gamma$. 

Orbital phases were computed using the most 
accurate quadratic ephemeris published by Parmar et al (2000). However, 
given the large number of cycles we estimate an accumulated uncertainty of 
0.03 orbits at the time of our observations and hence we decided to take 
the most updated $T_0$ from JvdK. This is defined 
by the superior conjunction of the X-ray pulsar and yields a final accuracy 
of 0.002 cycles. Binary phases are therefore computed from the ephemeris:

\begin{displaymath} 
T= 2450993.27968 + 0.232108785(50) E + 2.06(23) \times 10^{-11} E^2
\end{displaymath}
  
The orbital evolution of several lines is plotted in Fig. 2 in the 
form of trailed spectrograms. Balmer lines are dominated by deep 
absorptions, with velocity shifts of $\sim$ 800 km s$^{-1}$, 
maximum depth between phases 0.7-1.0 and strengthening with 
excitation energy. Some 
residual emission cores are nonetheless seen in H$\beta$ and H$\gamma$ 
at phase 0-0.1. The HeI lines are always seen in absorption with 
lower amplitudes ($\sim$ 200 km s$^{-1}$) and blue-to-red crossing 
velocities at phase 0 (e.g. see HeI $\lambda$4471). 
%These properties point towards a possible origin on the companion star. 
A radial velocity 
curve is presented in the top panel of figure 3, obtained after  
cross-correlating the spectra, coadded in 13 phase bins, with a gaussian 
of $FWHM=500$ km s$^{-1}$. As noted by Harlaftis et al. (1997; see also Jonker 
et al. 2002), the radial velocity curve is distorted around phase 0 because 
of possible contamination by another component. Therefore, we mask the data 
points within the phase interval 0.95-0.25 and perform a sinewave fit to the 
remaining to derive the following parameters: $\gamma=24 \pm 
28$ km s$^{-1}$, $V=229 \pm 37$ km  s$^{-1}$ and $\Phi_0=-0.07 \pm 0.02$, 
where $V$ is the velocity semi-amplitude and $\Phi_0$ the blue-to-red 
crossing. These numbers are in good agreement with Harlaftis et al. (1997) 
and the phase shift confirm
the claim by Jonker et al. (2002) that the radial velocity curve of HeI  
occurs earlier than expected from the pulse-timing ephemeris. 

The HeII $\lambda$4686 emission exhibits a classic double-peaked disc 
profile with an S-wave crossing from blue to red velocities at phase 
$\sim$ 0.8.  We postpone a detailed analysis of the radial velocity 
curve of HeII to Sect. 5.
On the other hand, the $\lambda$3813 line displays a 
single S-wave, significantly broader (FWHM $\sim 450$ km 
s$^{-1}$) than our resolution, and blue-to-red crossing also 
at phase 0.8. The S-wave pattern of this feature in our trailed spectra 
demonstrates that it is a real line and not just a spurious feature 
caused by the broadened Balmer absorptions at each side (Mason et al. 1982).
As noted by Charles et al. (1980), if it were HeII $\lambda$3813 there is no 
special reason why this line should be enhanced relative to others in the 
Pickering series. Alternatively, it could be the blue member of the OVI 
doublet $\lambda$3811/$\lambda$3834 and indeed we see another S-wave 
emission in our trailed spectra at $\sim \lambda$3834, although much weaker 
because of overlapping with the broad H9 absorption. Therefore, we identify 
this feature with OVI $\lambda$3811.4. We have also extracted the radial 
velocity curve of this line by cross-correlation of our 13 phase bins 
of co-added spectra with a gaussian of $FWHM=600$ km s$^{-1}$, and the 
results are presented in fig. 3 (middle panel). A sinewave fit yields 
$\gamma=46 \pm 24$ km s$^{-1}$, $V=370 \pm 30$ km s$^{-1}$ and 
$\Phi_0=0.82 \pm 0.01$. 
The phasing and large velocity amplitude suggest an origin related to the 
accretion disc/hot-spot region. 
  
The Bowen blend shows a complicated pattern of narrow  
(FWHM $\sim 100$ km s$^{-1}$) S-waves, barely detectable in 
our limited signal-to-noise ratio individual spectra, superposed on the broad 
disc emission. These narrow components are identified with 
the Bowen fluorescent components of NIII 
4634.12/4640.64\AA~ and CIII 4647.4/4650.1\AA~ and have been shown to arise 
from the irradiated hemisphere of the donor in Sco X-1 
(see SC). 
We have attempted to extract radial velocities of these sharp Bowen 
fluorescent components using a multigaussian fit, as explained in SC, but 
the method failed due to the weakness of these features 
in the individual spectra. However, in Section 6 we will show how we 
can exploit the powerful Doppler Tomography technique (Marsh 2001) 
to derive the radial velocity curve of the $\lambda$4640.64 NIII line, which
is found to trace the motion of the donor star. 
The reliability of this technique was already demonstrated 
in the case of Sco X-1; radial velocities derived from Doppler maps were in 
excellent agreement with the results from Gaussian fitting (SC).
%We have measured its velocity 
%by cross-correlation with a gaussian of fixed $\sigma=190$ km s$^{-1}$. 
%A sinewave fit to the radial velocity curve is shown in Fig. 8 and it yields 
%a velocity semiamplitude of 390 $\pm$ 23 km s$^{-1}$.

\section{Light Curves}

In the top left panel of Figure 4 we show the continuum light curve, 
extracted from integrating the individual fluxed spectra in the wavelength 
range $\lambda\lambda$4000-4600. Solid circles correspond to the night of 10 
July and open circles to 11 July.  In addition to the 
eclipse of the disc by the companion star, we also note a post-eclipse 
hump centered at orbital phase 0.25-0.3  which has been interpreted as the 
visibility of the irradiated inner disc rim or hot-spot hump 
(e.g. Mason \& C\'ordova 1982). We note that this is when the OVI flux (top 
right panel) is also at a maximum, suggesting that it too is located in this 
region.
%The light curve morphology is reminiscent of some novalikes of the 
%SW Sex class (e.g. WX Ari, Rodr\'\i{}guez-Gil et al. 2000)
% and is different to quiescent dwarf novae, dominated by the 
%visibility of the hot-spot which produces a pre-eclipse hump at 
%phase 0.8 (e.g. Z Cha, ref). 
%In order to search for colour variability due to X-ray heating 
%effects and the eclipse of the accretion disc, we have extracted fluxes 
%in two spectral bands at $\lambda$4040 and $\lambda$4250, with 100 \AA width. 
%These regions have been selected because they are free from emission lines. 
%The bottom panel of Fig. 2 displays the flux ratio between these two bands 
%but no colour modulation with orbital phase is clearly observable.

We have also obtained integrated light curves of the main emission/absorption 
lines and these are also presented in Fig. 4. The light 
curve of HeII $\lambda$4686 shows a narrow 
eclipse at phase 0.2 and maximum emission around phase 0.75, suggesting that 
the bulk of the emission might be related to the visibility of 
the stream/disc impact region (i.e. equivalent to the classical hot spot in 
a CV). The Bowen blend exhibits a sinewave modulation 
peaking at phase 0.5, which suggests a likely origin on the irradiated 
hemisphere of the donor star. We note that the phases of maximum emission of 
both HeII and Bowen are the same as observed in the neutron star transient  
XTE J2123-058 during its outburst phase (Hynes et al. 2001).

The OVI $\lambda$3811 emission line shows a similar light curve 
to the Bowen blend, but leading in phase by $\sim$ 0.15 orbits.  
The minimum flux occurs at phase 0.8, which also coincides with the 
blue-to-red crossing (see Fig. 3). The combination of the light curve and 
velocity information strongly suggests that OVI $\lambda$3811 
is associated with the irradiated {\it inner} rim of the hot-spot bulge, which 
is self-obscured by the outer disc rim at phase $\simeq $ 0.8. The amplitude 
of the modulation is a factor of two larger than in the Bowen blend, and also 
the flux 
drops to zero at minimum, indicating that the entire line is produced at 
the irradiated inner bulge, with no significant contribution from other 
regions (e.g. axisymmetric disc or companion star).
None of the emission lines exhibit a sharp eclipse feature at phase 0 
and hence they must arise in a vertically extended structure, such as a 
disc atmosphere or corona.   

The Balmer lines show intricate light curves in absorption.
They exhibit two minima, at phase 0.25 and 0.8, increasing  
depth with excitation energy i.e. deeper for H8 than for 
H$\beta$. The trailed spectra of H$\gamma$ in figure 2 clearly show 
that the phase 0.25 dip is produced by a sharp absorption whereas the 0.8 
minimum is caused by a broader feature moving redwards from phase 0.5 to 1.0. 
The phase 0.8 absorption is consistent with obscuration of the irradiated 
inner disc by a high disc rim extending from phase 0.5 to 0.9.    
The light curve of HeI $\lambda$4471, also in absorption, appears to modulate 
only weakly with orbital phase, reaching  maximum strength around phase 0.6. 

\section{The Systemic Velocity}

Before discussing the location of the different emission sites we need to 
estimate the systemic velocity from these data. So far there is no 
clear determination, with only tentative values based on 
radial velocity curves of the HeI lines: 19 $\pm$ 19 km s$^{-1}$ (Harlaftis et 
al. 1997) and 67 $\pm$ 15 km s$^{-1}$ (Jonker et al. 2002).  
On the other hand, we notice that the strong HeII $\lambda$4686 emission 
is blueshifted in our sum spectrum and a simple Gaussian fit yields a 
centroid position at -50 km s$^{-1}$. The double-peaked profile 
indicates it comes from the accretion disc and hence it will likely provide 
the most reliable $\gamma$ determination. 

In an attempt to estimate both the systemic velocity and the velocity 
semi-amplitude of the compact object $K_1$, we have applied the 
double-Gaussian technique to HeII $\lambda$4686 (Schneider \& Young 1980). 
We employed a two Gaussian band-pass with $FWHM=100$ km s$^{-1}$ and 
Gaussian separations $a=400-1300$ km s$^{-1}$ in steps of 100 km 
s$^{-1}$. The $K_1$ velocity drops dramatically from 230 to 
$\simeq$ 60 km s$^{-1}$ as we move from the line core to the wings, 
whereas the systemic velocity yields consistent values in the range 
-30 -- -60 km s$^{-1}$. We find that the velocity points start to be 
corrupted by continuum noise for $a > 900$ km s$^{-1}$ and a 
sinewave fit (with the period fixed) to this velocity curve 
gives $K_1=63 \pm 3$ km s$^{-1}$, $\phi_{0}=0.58 \pm 0.01$ and $\gamma=-43 
\pm 4$ km s$^{-1}$ (see bottom panel in figure 3). On the other hand, 
the radial velocity curve of the compact object is well determined 
through the pulse-timing analysis and yields $K_1=94.5 \pm 0.5$ km 
s$^{-1}$ (represented by a dotted 
curve in fig. 3). The small difference in phasing (0.08 orbits) and 
$K_1$ velocity indicates that even the line wings are contaminated by 
inhomogeneities in the disc. Nevertheless, the systemic velocity 
appears quite stable and hereafter in this paper 
we will adopt $\gamma=-43$ km s$^{-1}$ from the fit using $a=900$ km 
s$^{-1}$. This is subject to systematic uncertainties, but further 
confirmation for a systemic velocity close to -43 will be presented in 
the next section. 
At this point we note that the radial velocity curve of OVI    
implies a positive systemic velocity of $\sim$ 43 km s$^{-1}$ (see sect.  
3). This is due to its formation in gas located on the irradiated inner face
of the hot spot which probably has a local vertical velocity  
component after the gas stream collision with the disc outer edge (see 
fig. 5).

\section{Doppler Mapping}

Armed with this value of $\gamma$, we have computed Doppler maps 
of the principal lines seen in our spectra and these are presented in Fig. 5. 
All the spectra were rectified, by 
subtracting a low-order spline fit to the continuum, and rebinned into 
a uniform velocity scale of 15 km $s^{-1}$. Doppler Tomography combines 
the orbitally-resolved line profiles to reconstruct the brightness 
distribution of the system in velocity space by maximizing the entropy 
of the final image (see Marsh 2001 for details). 
Since Doppler Tomography only deals with emission features, in order 
to map the absorption line HeI $\lambda$4471 we first had to invert the 
spectra. 
%In all the maps we have adopted $\gamma=-43$ km s$^{-1}$.

The emission line distribution of HeII $\lambda$4686 shows the classic 
ring-like distribution of an accretion disc. The distribution is slightly 
asymmetric, with maximum emissivity over an extended arc peaking at phase 
$\sim$ 0.8.
%along the negative velocities 
%quadrant. This is a signature of SW Sex stars and it has also been 
%detected in the HeII Doppler map of the neutron star transient XTE J2123-058 
%during outburst (Hynes et al. 2000). It has been suggested that this emission
%arises from disc overflow material which speeds down and cools after 
%colliding with the outer edge of the disc or, alternatively, from collision 
%of gas blobs which are propelled by magnetic fields in the disc or the 
%magnetosphere of the compact star (Horne 1999, Wynn, King \& Horne 1997). 
In order to help interpret the maps we have overplotted the Roche lobe of 
the companion star and the gas stream trajectory for a typical neutron star 
mass of $M_1=1.4 M_{\odot}$ and hence $q=M_2/M_1 = K_1/K_2 = 0.29$. We have 
also marked the center of mass and compact object's position with crosses.  

We used the HeII Doppler map to verify the adopted systemic velocity. The 
$\chi^2$ value of the map was calculated for a range of $\gamma$'s, and the 
best fit in terms of mimimal $\chi^2$ was achieved for $\gamma=-44 \pm 5$  
km s$^{-1}$. We also searched for the optimal center of symmetry of the HeII 
map as a means to estimate $K_1$ (see SC). We found that 
the disc emission was subtracted optimally by assuming a $K_1$ of 97 
$\pm$ 5 km s$^{-1}$. We did not include the region at negative $V_x$ 
velocities, since this part of the map is severely affected by the bright 
spot asymmetries. This value compares favorably with the radial velocity 
amplitude of the primary as determined from the X-ray pulsations (JvdK).

The Doppler image of OVI $\lambda$3811 shows no evidence for the accretion 
disc but a very compact emission knot which lies between the predicted 
gas stream velocity and the keplerian velocity of the disc along the gas 
stream path. The OVI emission probably arises from irradiated turbulent gas  
around the hot spot which, after the shock, shares the velocity of the 
gas stream and the accretion disc. This is identical to the Doppler maps 
of HeII $\lambda$4686 and HeI $\lambda$4471 computed for the cataclysmic 
variable U Gem (Marsh et al. 1990). 

%The tomograms of the Balmer lines show extended structures in the 
%same velocity region with maximum absorption at higher velocities and 
%they are probably produced by absorption of disc light by a large cloud 
%of overflowed/propelled expanding gas. 
 
The Doppler analysis of the Bowen blend is complicated by the merging of 
many CIII, NIII and OII transitions (e.g. McClintock, Canizares \& 
Tarter 1975). The most prominent component is NIII 4640.64\AA~ for which 
we have also computed a tomogram. We clearly 
detect a sharp spot along the $V_y$ axis and we have inferred 
its centroid position at $V_x \sim -12 \pm 8$ km s$^{-1}$ and $V_y \sim 300 
\pm 8$ km s$^{-1}$ through a two-dimensional gaussian fit.
The spot is effectively on the $V_y$ axis given our phase uncertainties. 
The phasing and velocity amplitude proves that this emission is produced 
on the irradiated front side of the donor. 
We have experimented with computing maps for a set of $\gamma$-velocities in 
the range 0 -- -80 km s$^{-1}$ and we find that only $\gamma$ values in the
range -40 -- -60 km s$^{-1}$ provide well focused NIII spots. The remaining 
$\gamma$-velocity maps produce more extended and blurred spots which further 
confirm our systemic velocity determination.
In Fig. 6 we compare the Bowen blend average of X1822-371 and Sco X-1 
(from SC), in the rest frame of the donor stars. 
The prominent NIII $\lambda$4640 emission and the CIII blend at 
$\lambda$4647/50 are clearly identified in X1822-371, whereas the NIII 4634 
is diluted in the much noisier spectrum of X1822-371.

Finally, we also present the tomogram of HeI $\lambda$4471 absorption, which 
very much resembles the map of NIII 4640.64\AA~
but the spot seems to be skewed to the leading side of the donor or 
on the gas stream, at 0.1 $R_{\rm L1}$ 
from the $L_1$ point. A two-dimensional gaussian fit to the bright spot yields 
a centroid position at $V_x = -74 \pm 8$ km s$^{-1}$ and $V_y \sim 224 
\pm 8 $km s$^{-1}$  which places it exactly over the gas stream 
trajectory. 

\section{Discussion}

We have succesfully detected the companion star in X1822-371 through 
Doppler imaging of the fluorescent NIII $\lambda$4640 emission. The 
emission is exactly in antiphase with the radial velocity curve of the 
pulsar and hence no asymmetric irradiation of the companion's Roche 
lobe needs to be invoked. 
On the other hand, the distribution of HeI $\lambda$4471 absorption 
is concentrated near the $L_1$ point but leading in phase. We find a 
phase shift of $\simeq$ -0.05 with respect to the position of the 
NIII $\lambda$4640 spot and the pulse-timing ephemeris 
which places the HeI absorber on the leading side of the companion's Roche lobe or, 
more likely, over the gas stream. This solves the 
problem raised by Jonker et al. (2002) who, following Harlaftis et al. (1997), 
assumed that the HeI absorption arises on the irradiated hemisphere of the  
companion.  

The Doppler map of NIII $\lambda$4640 yields a bright 
spot at $V_y = 300 \pm 8 $km s$^{-1}$. Since this spot is produced on the 
inner hemisphere of the companion star through irradiation, it gives  
a lower limit to $K_2$, the velocity semi-amplitude of the 
companion's center of mass. 
The lower limit to $K_2$ can then be combined with the very 
accurate determination of the inclination ($i=82.5^{\circ} \pm 
1.5^{\circ}$) and the radial velocity curve of the pulsar to derive 
robust lower limits to the masses of the neutron star and companion of 
$M_1 \ge 1.14 \pm 0.06 M_{\odot}$ and  $M_2 \ge 0.36 \pm 0.02 M_{\odot}$, 
respectively.

An upper limit to $K_2$ can be 
obtained by using the K-correction equation (Wade \& Horne 1988): 

\begin{displaymath} 
K_2 = K_{\rm em} / \left( 1 - f~q^{1/3} (1+q)^{2/3} \right) ,
\end{displaymath}

\noindent where $K_{\rm em}$ is the velocity semi-amplitude of the 
irradiated-induced 
emission, $q=M_2/M_1$ and $f\leq 1$ is the displacement of the reprocessing 
site from the center of mass of the star in units of the companion's star 
radius $R_2$. Furthermore we assume that the NIII line (with $K_{\rm em} = 
300$ km s$^{-1}$) is produced at 
the $L_1$ point i.e. $f=1$. The equation then yields $K_2 \le 437$ km 
s$^{-1}$, where we have used $q=K_1/K_2$ and $K_1=94.5$ km s$^{-1}$. 
This upper limit would imply $M_1 \le 3.1 M_{\odot}$ which coincides 
with the maximum neutron star mass possible with the stiffest equation 
of state of condensed matter. In addition, since 
$K_2$ must be between 300--437 km s$^{-1}$ and 
we know $K_1$, this translates to firm limits to the mass ratio of 
$ 0.216 < q < 0.315$.

For the case of a canonical neutron star mass of 1.4 
$M_{\odot}$ we would expect $K_2 = 324$ km s$^{-1}$ and $q=0.29$. This 
would imply a K-correction for NIII of $K_{\rm em}/K_2 = 0.93$ which is 
comensurate with the value seen for the cataclysmic variable 
HU Aqr which has a similar mass ratio (Schwope, Mantel \& Horne
1997). On the other hand, JvdK have shown 
that the neutron star in X1822-371 is undergoing an accretion spin-up 
phase and therefore we might expect to find $M_1 > 1.4 
M_{\odot}$. Tighter constraints on the system parameters may have 
important implications for our knowledge of the equation of state of 
nuclear matter (e.g. Cook, Shapiro \& Teukolsky 1994). 

The combination of light curves and velocity information from trailed 
spectra/Doppler maps has enabled us to locate the different regions 
responsible for the main spectral lines in the binary. 
The Balmer lines (from H$\beta$ to the Balmer break) are embedded in 
deep absorptions whose phasing and large velocity shifts suggest 
they are produced in vertically extended cool gas associated with the 
hot-spot/post-shock region. Simulations of stream-disc interaction 
with inefficient cooling (Armitage \& Livio 1998) produce a 
``halo'' of material above the disc and a bulge along the disc rim 
which can obscure the bright inner disc. 
This behaviour is mainly observed in 
SW Sex stars (e.g. Groot, Rutten \& van Paradijs 2001, Rodr\'\i{}guez-Gil, 
Mart\'\i{}nez-Pais \& Casares 1999) where (possible) 
magnetic white dwarfs are fed by disc 
overflowing material. SW Sex behaviour has also been recognized in the 
neutron star transient XTE J2123-058 during outburst (Hynes et al. 2001). 
A key ingredient of this behaviour seems to depend on large accretion rates, 
although an inclination selection effect may be needed to account for the 
lack of orbitally modulated absorptions in other persistent LMXBs (e.g. 
Sco X-1). 
The OVI $\lambda$3811 emission arises from the post-shock region between 
the gas stream and the disc but it has maximum flux at phase 0.25, rather 
than 0.8. Therefore, it must be triggered by irradiation of the 
inner hot-spot bulge region i.e. further evidence for a vertical 
thickening of the disc at the splash point position.   

After our work on Sco X-1 (SC) and GX339-4 (Hynes et 
al. 2002) this paper further establishes the importance of the fluorescent 
Bowen as excellent tracer of the companion's motion 
in active LMXBs and Soft X-ray Transients in outburst. This new technique 
opens a new window on the detection of the, otherwise, radiatively 
overwhelmed companion stars and thereby allows follow-up dynamical studies 
of X-ray binaries.

\acknowledgments

DS acknowledges the support of a PPARC Postdoctoral Fellowship and a
Smithsonian Astrophysical Observatory Clay Fellowship.  RIH and PAC
acknowledge support from grant F/00-180/A from the Leverhulme Trust.  RIH
is currently funded from NASA through Hubble Fellowship grant
\#HF-01150.01-A awarded by the Space Telescope Science Institute, which is
operated by the Association of Universities for Research in Astronomy,
Inc., for NASA, under contract NAS 5-26555.
MOLLY and DOPPLER software developed by T.R. Marsh is gratefully 
acknowledge. Partly based on data collected at the European Southern 
Observatory, La Silla, Chile.

\clearpage


\begin{figure}
\plotone{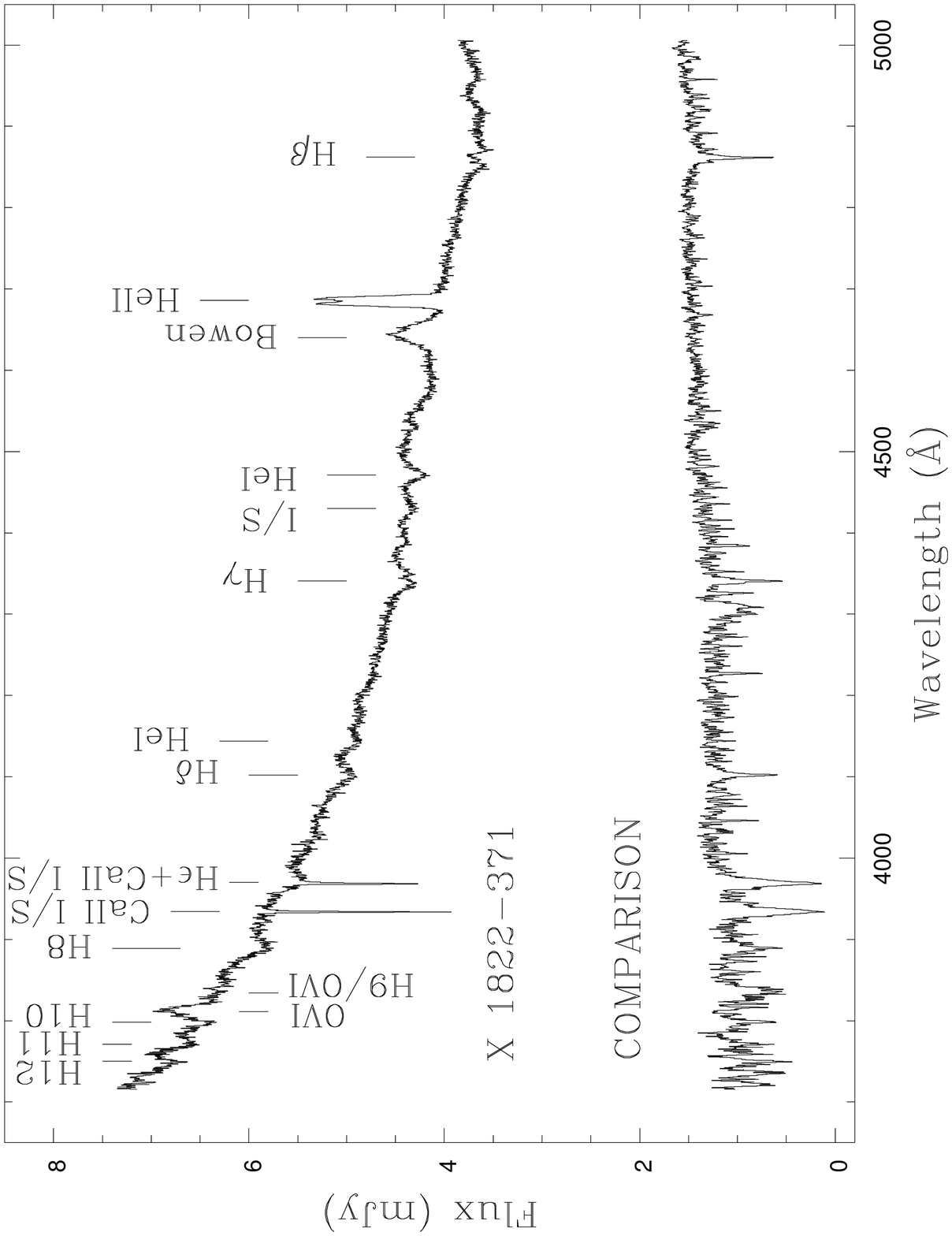}
\caption{Average optical spectrum of X1822-371 and 
the comparison star in the slit. Main emission and absorption features are
indicated. \label{fig1}}
\end{figure}

\clearpage 

\begin{figure}
\plotone{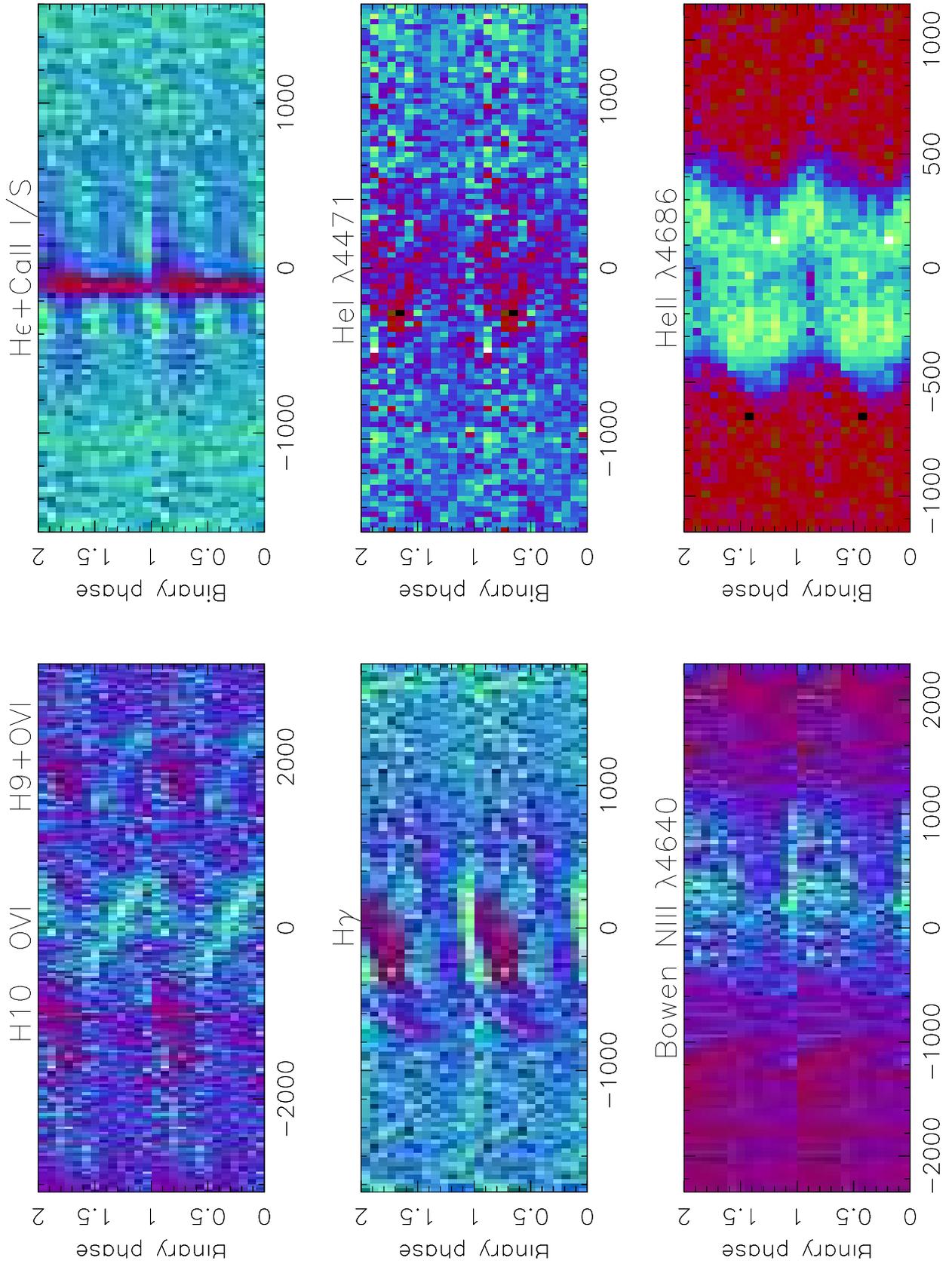}
\caption{Trailed spectrogram of selected lines folded in 13 phase 
bins.\label{fig2}}
\end{figure}

\clearpage 

\begin{figure}
\plotone{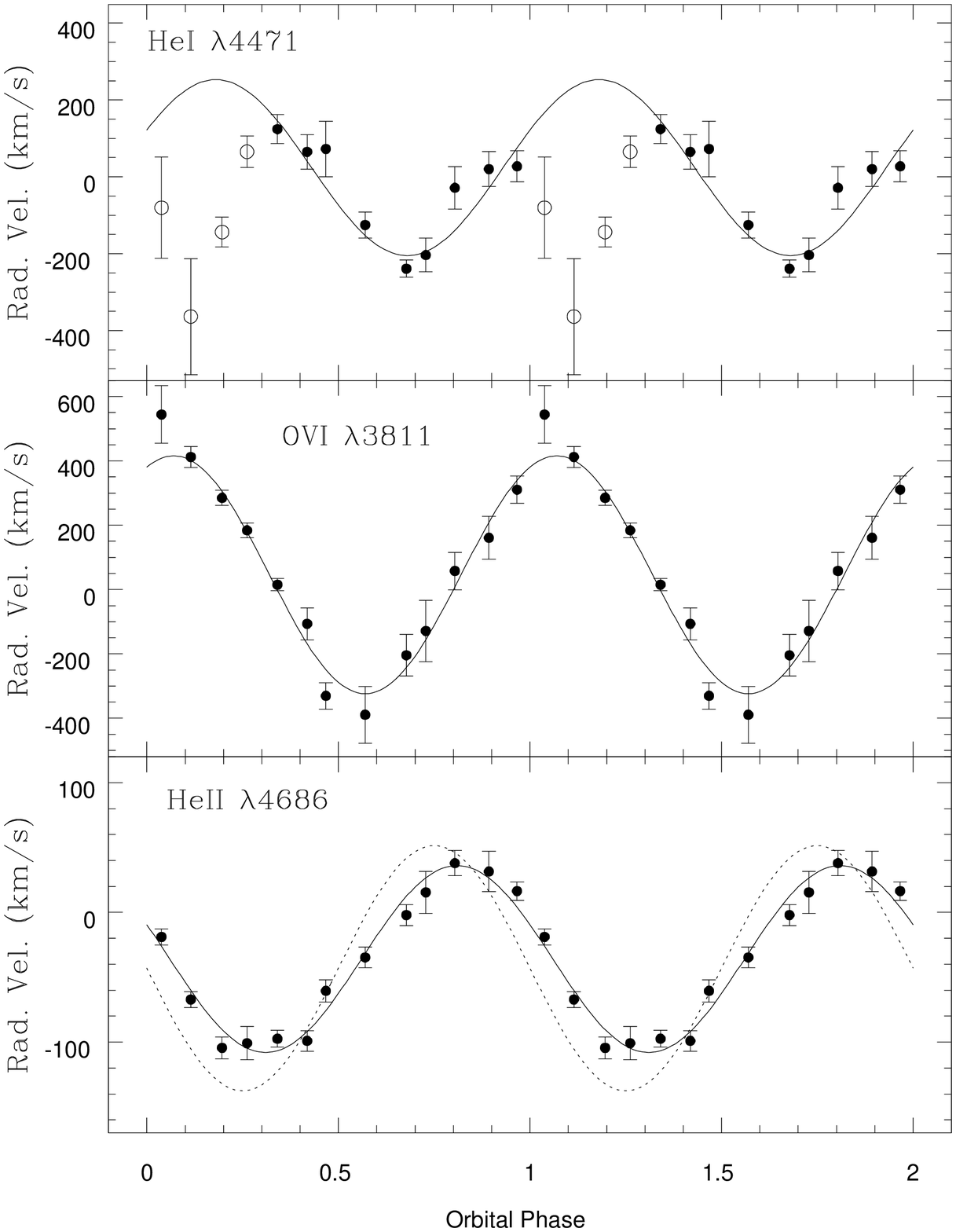}
\caption{Radial velocity curves of HeI $\lambda$4471 
(top panel), OVI $\lambda3811$ (middle panel) and the wings of HeII 
$\lambda$4686 (bottom panel). The latter was obtained by cross-correlation 
with a double gaussian template of $FWHM=100$ km s$^{-1}$ and separation 
$a=900$ km s$^{-1}$ (see Sect. 5 for details). Best sinewave fits are 
shown, after masking a few outliers, marked with open circles, for HeI 
$\lambda$4471 (see text for details).   
The dotted line displays the radial velocity curve of the neutron star 
using the pulse-timing ephemeris from JvdK.\label{fig3}}
\end{figure}

\clearpage 

\begin{figure}
\plotone{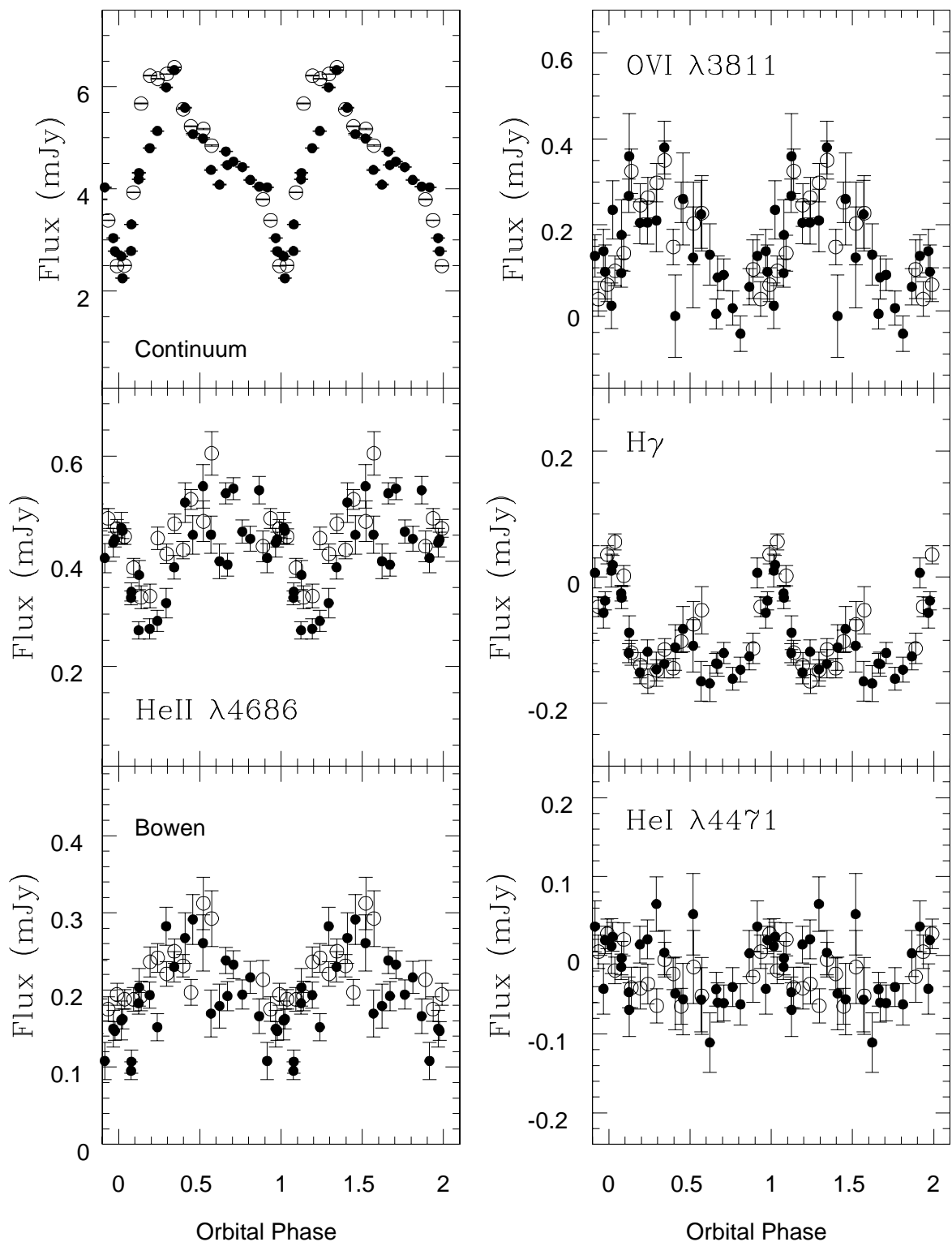}
\caption{Light curves of the continuum ($\lambda\lambda$4000-4600) 
and selected emission/absorption 
lines. Solid and open circles correspond to the nights 
of 10 and 11 June, respectively.\label{fig4}}
\end{figure}

\clearpage 

\begin{figure}
\plotone{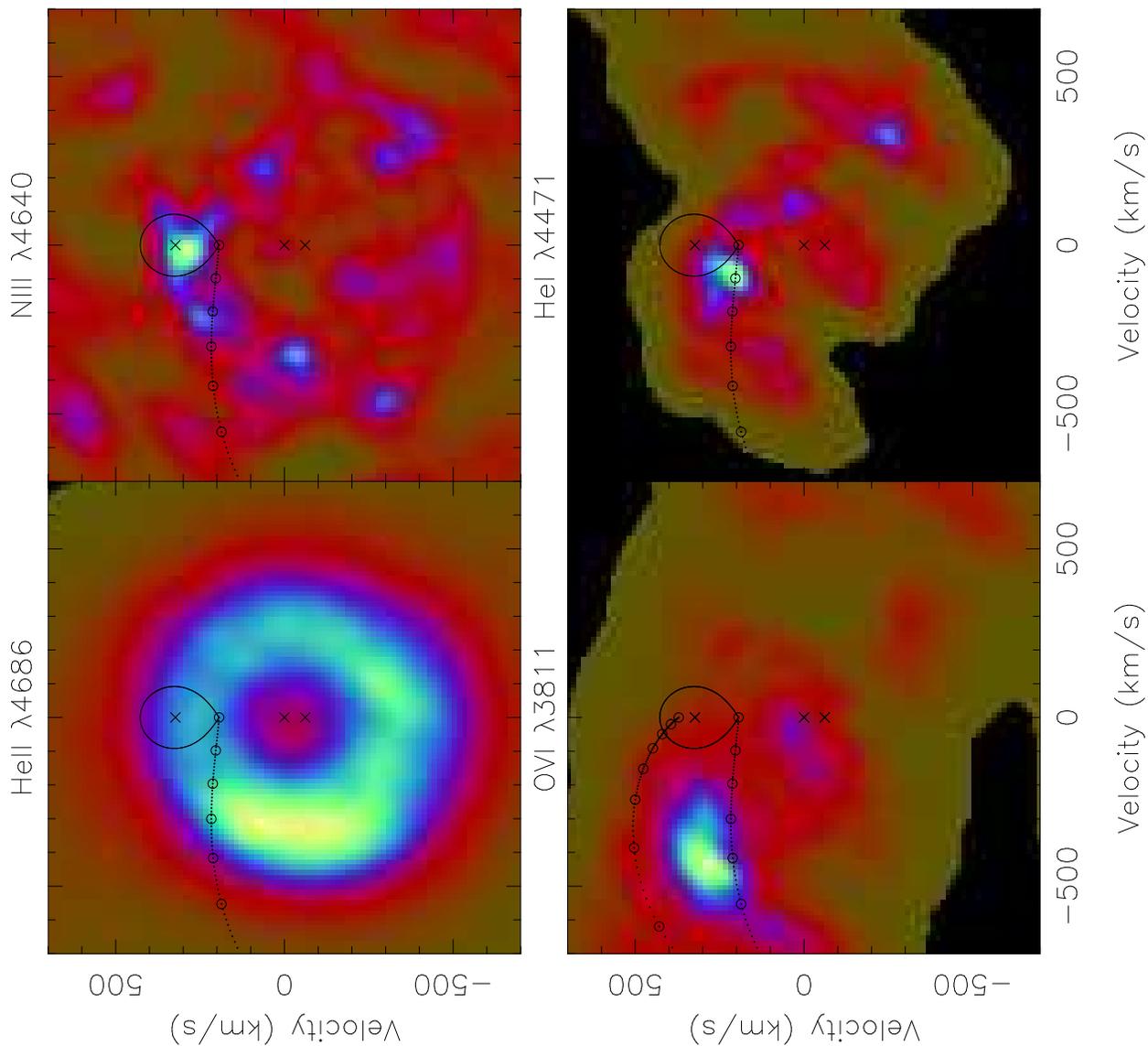}
\caption{Doppler maps of several spectral features. We have 
overplotted the Roche lobe of the companion star, the gas stream trajectory 
and the compact object's position for $q=0.29$, which assumes $M_1=1.4
M_{\odot}$ (i.e. $K_2=324$ km s$^{-1}$). In the map of OVI we also plot 
the keplerian velocity of the disc along the gas stream path. All the   
images were computed for a systemic velocity $\gamma = -43$ km s$^{-1}$.
\label{fig5}}
\end{figure}

\clearpage 

\begin{figure}
\plotone{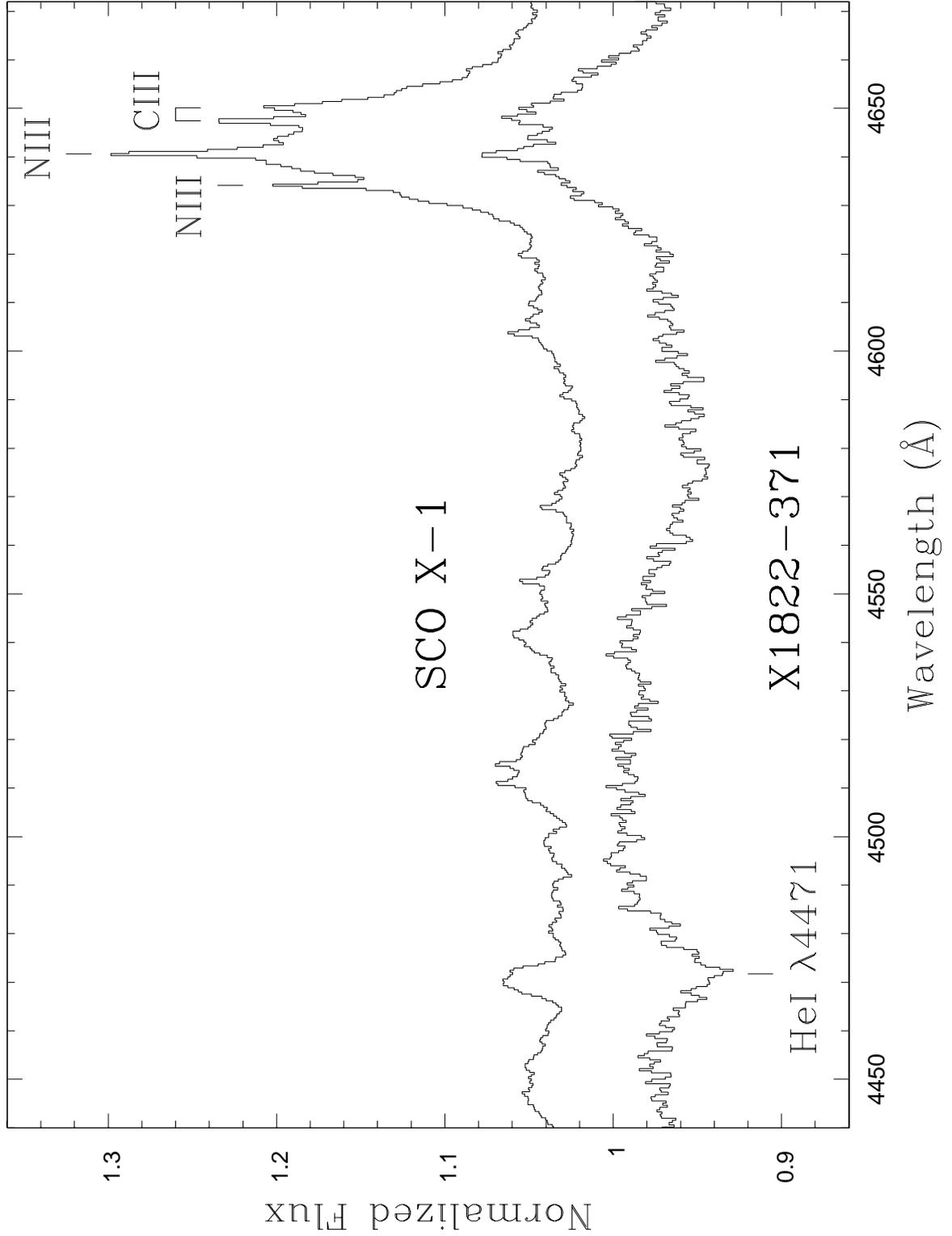}
\caption{Doppler corrected averaged spectra of Sco X-1 and 
X1822-371, in the rest frame of the companion. In the latter case we have 
used $\gamma= -43$ km s$^{-1}$ and $K_2=300$ km s$^{-1}$.\label{fig6}}
\end{figure}

%% Tables should be submitted one per page, so put a \clearpage before

\clearpage

\begin{deluxetable}{lccccc}
%\tabletypesize{\scriptsize}
\tablecaption{Log of the observations . \label{tbl-1}}
\tablewidth{0pt}
\tablehead{
\colhead{Date} & \colhead{Telescope}   & \colhead{Exposure Time} &
\colhead{Wav. Range} & \colhead{Dispersion}  \\
& \colhead{(seconds)} & \colhead{\AA} & \colhead{\AA~$pix^{-1}$)} 
}
\startdata
10 Jun 02 &  AAT &  24x900  & 3650-5100 & 0.45  \\
11 Jun 02 &   ,, &  14x900  &	 ,,	&  ,,	\\
   ,,     &  NTT &  1x1200  & 4400-5150 & 0.40  \\
   ,,     &   ,, &   4x600  &	 ,,	&  ,,	\\
\enddata
\end{deluxetable}


\begin{thebibliography}{}

%\bibitem[Armitage \& Livio 1996]{armitage96}Armitage P.J., Livio, M. 1996, \apj, 
%470, 1024
\bibitem[Armitage \& Livio 1998]{armitage98}Armitage P.J., Livio, M. 1998, \apj, 
493, 898
\bibitem[Charles et al. 1980]{charles80}Charles P., Thorstensen J.R., Barr P. 1980, \apj, 241, 1148
\bibitem[Cook et al. 1994]{cook94}Cook G.B., Shapiro S.L., Teukolsky S.A. 1994, \apj, 424, 823
\bibitem[Groot et al. 2001]{groot01}Groot P.J., Rutten R.G.M., van Paradijs J. 2001, \aa, 368, 183
\bibitem[Harlaftis et al. 1997]{harlaftis97}Harlaftis E.T., Charles P.A., Horne K. 1997, \mnras, 285, 673
\bibitem[Heinz \& Novak 2001]{heinz01}Heinz S, Nowak M.A. 2001, \mnras, 320, 249 
\bibitem[Hellier \& Mason 1989]{hellier89}Hellier C., Mason K.O. 1989, \mnras, 239, 715
\bibitem[Horne 1986]{horne86}Horne K. 1986, \pasp, 98, 609
\bibitem[Hynes et al. 2001]{hynes01}Hynes R. I., Charles P. A., Haswell C. A., Casares, J., Zurita, C., 
Serra-Ricart M. 2001, \mnras, 324, 180
\bibitem[Hynes et al. 2003]{hynes03}Hynes R.I., Steeghs D., Casares J., P.A. Charles, 
O'Brien K. 2003, \apj Letters, in press, astro-ph/0301127
\bibitem[Jonker \& van der Klis 2001]{jonker01}Jonker P.G., van der Klis M. 2001, 
\apj, 553, L43
\bibitem[Jonker et al. 2002]{jonker02}Jonker P.G., van der Klis M., Groot P.J. 
2002, \apj, astro-ph/0210506
\bibitem[Marsh 2001]{marsh01}Marsh T.R. 2001, in Astrotomography, Indirect Imaging Methods in 
Observational Astronomy, eds. H.M.J. Boffin, D. Steeghs and J. Cuypers 
(Lecture Notes in  Physics) vol. 573, p.1
\bibitem[Marsh et al. 1990]{marsh90}Marsh T.R., Horne K., Schlegel E.M., Honeycutt R.K., 
Kaitchuck R.H. 1990, \apj, 364, 637  
\bibitem[Mason \& C\'ordova 1982]{mason82}Mason K. O., C\'ordova F.A. 1982, \apj, 262, 253
\bibitem[Mason et al. 1982]{mason82b}Mason K. O., Murdin P.G., Tuohy I.R., Seitzer P., 
Branduardi-Raymont G. 1982, \mnras, 200, 793
\bibitem[McClintock et al. 1975]{mcclintock75}McClintock J.E., Canizares C.R., 
Tarter C.B. 1975, \apj, 198, 641
\bibitem[Parmar et al. 2000]{parmar00}Parmar A.N. et al. 2000, \aa, 356, 175
\bibitem[Rodr\'\i{}guez-Gil et al. 1999]{rodriguez99}Rodr\'\i{}guez-Gil P., Casares J., Mart\'\i{}nez-Pais I.G. 
1999, \mnras, 305, 661
\bibitem[Schwope et al. 1997]{schwope97}Schwope A. D., Mantel, K.-H., Horne K. 1997, \aa, 319, 894
\bibitem[Steeghs \& Casares 2002]{steeghs02}Steeghs D., Casares J. 2002, \apj, 568, 273
\bibitem[Schneider \& Young 1980]{schenider80}Schneider D.P., Young P. 1980, 
\apj, 238, 946
\bibitem[]{van1}van Paradijs J., McClintock J.E. 1995, in X-Ray Binaries, 
eds. W.H.G. Lewin, J. van Paradijs and E.P.J. van den Heuvel (CUP 26,
Cambridge), p58 
\bibitem[Wade \& Horne 1988]{wade88}Wade R.A., Horne, K. 1988, \apj, 324, 411
\bibitem[White \& Holt]{white82}White N.A., Holt S.S. 1982, \apj, 257, 318 

\end{thebibliography}
\end{document}